\documentclass[12pt,twocolumn]{aastex631}
\usepackage{verbatim, graphicx, physics, hyperref}
\usepackage{rotating}

\graphicspath{{./}{figures/}}

\begin{document}


\title{Preparing for the Early eVolution Explorer: The Impact of Flare Temperature on Ozone Column Depth in Earth-Like Atmospheres}

\author[0000-0003-2273-8324]{Jaime S. Crouse}
\affiliation{John Hopkins University, 3400 N. Charles Street, Baltimore, MD 21218, USA}
\affiliation{NASA Goddard Space Flight Center, 8800 Greenbelt Road, Greenbelt, MD 20771}
\email{jcrouse3@jh.edu}

\author[0000-0002-0413-3308]{Nicholas F. Wogan}
\affiliation{SETI Institute, Mountain View, CA 94043, USA}
\affiliation{NASA Ames Research Center, Moffett Field, CA 94035, USA}

\author[0000-0002-0583-0949]{Ward S. Howard}
\affiliation{NASA Hubble Fellowship Program Sagan Fellow}
\affiliation{Department of Astrophysical and Planetary Sciences, University of Colorado, 2000 Colorado Avenue, Boulder, CO 80309, USA}

\author[0000-0001-7891-8143]{Meredith A. MacGregor}
\affiliation{John Hopkins University, 3400 N. Charles Street, Baltimore, MD 21218, USA}

\author[0000-0001-5455-6678]{Guadalupe Tovar Mendoza}
\affiliation{John Hopkins University, 3400 N. Charles Street, Baltimore, MD 21218, USA}

\author[0000-0002-0746-1980]{Jacob Lustig-Yaeger}
\affiliation{Johns Hopkins Applied Physics Laboratory, 11100 Johns Hopkins Rd, Laurel, MD 20723, USA}

\author[0000-0002-7260-5821]{Evgenya L. Shkolnik}
\affiliation{School of Earth and Space Exploration, Arizona State University, 781 Terrace Mall, Tempe, AZ, USA, 85287}

\begin{abstract}

Atmospheric photochemical models incorporating the impacts of stellar flares often assume a $\sim$9,000~K spectrum at ultraviolet-optical wavelengths. Recent multiwavelength observations, however, reveal a more complex picture with temperature measurements spanning 4,000--40,000~K, although the occurrence rates for flares with different temperatures remain unknown. Here, we model the evolution of a Proterozoic Earth-like world with 0.01~bar of O$_2$ under repeated flaring to identify the impact of flare effective temperatures.  We explore four scenarios -- two host star types (K2V and M2.5V) and two flare temperatures (9,000 K and 19,000 K) -- selected to bound the potential parameter space. The hotter flares have a larger impact on O$_3$ photochemistry for both stellar types. M-star planetary atmospheres are more volatile and exhibit rapid changes in their O$_3$ production and destruction rates. Meanwhile, K-star planetary atmospheres are more stable and are only impacted by the hottest flares, proving advantageous for biosignature searches. We simulate 0.2--1.0~$\mu$m reflected light spectra for all four scenarios, and find that 19,000~K flares can result in either production or destruction of O$_3$ depending on the host star spectral type increasing the 0.2~$\mu$m feature by $\sim$2$\times$ for the K2V star but decreasing it by 50\% for the M2.5V star.  Future missions such as the EVE SMEX mission concept will provide robust flare temperature constraints for young FGKM stars, which will serve as inputs to improve photochemical models to inform future HWO observations. 

\end{abstract}


\section{Introduction}
\label{sec:intro}


Terrestrial planets orbiting in the `habitable zone,' the range of distances from the host star where an Earth-like planet’s surface temperature allows for the presence of liquid water \citep{Kopparapu:2013}, are generally considered our best chance to search for life in the Universe and are therefore the targets of upcoming missions to characterize exoplanet atmospheres and search for biosignatures such as molecular oxygen and ozone. We know of approximately 70 potentially rocky planets with $0.5 < R_p < 2.5$~R$_\oplus$ or $0.1 < M_p <10$~M$_\oplus$ in the habitable zones of nearby stars and therefore accessible to atmospheric characterization \citep{Hill:2023,Bohl:2026}. However, the majority ($>70\%$) of these potentially habitable planets orbit M-type stars \citep{Henry:2006} -- stars much smaller and cooler than our Sun that are well-known for their frequent outbursts of radiation or flares \citep{Schneider:2018,France:2016}. Photochemical modeling of Archean planetary atmospheres subject to the time-averaged (i.e., out-of-flare) emission of M4--M8V host stars observe O$_3$ production that causes a $\sim$50\% weakening of the 0.2~$\mu$m Hartley band in reflected light that could be misunderstood as evidence for low levels of biogenic O$_2$ \citep{Davis:2026}. Beyond M-stars, most FGK stars have been shown to exhibit higher levels of stellar activity than the Sun \citep{Reinhold:2020}. In fact, many solar analogs are regularly observed to emit superflares \citep{Shibayama:2013, Notsu:2019}, events with energies $\geq10^{33}$ erg that are 10-10000$\times$ more energetic than the largest solar flares ever detected \citep{Vadilyev:2024}. Frequent stellar flaring is expected to impact the potential for candidate habitable worlds to host long-lived, complex life via thermal escape and photochemical processes (e.g. \citealt{Tilley:2019, doAmaral:2022, Howard:2025L, Pass:2025, Mamonova:2026}).

Most previous analyses of the role of stellar flares in atmospheric photochemistry have largely focused on isolated superflare events from M stars.  These models generally use initial atmospheric compositions similar to modern Earth (O$_2$ $+$ N$_2$) or earlier stages of Earth’s evolution (varying levels of CO$_2$, N$_2$, and O$_2$) to determine the influence of flares on the thickness of the ozone column and the UV environment at the surface \citep{Segura:2010,Omalley:2018,Tilley:2019,Howard:2018,Segura:2007,Estrela:2018}, and are agnostic to the source of O$_2$ in the atmospheres, which could be of either biotic or abiotic origin. Several avenues exist to generate O$_2$-rich atmospheres on abiotic worlds. Abiotic oxygen can be produced by hydrogen loss from habitable zone planets \citep{Wordsworth:2013,Wordsworth:2014} with N$_2$-poor, CO$_2$-rich atmospheres. Photochemistry can also produce stable O$_2$ and O$_3$ abundances from CO$_2$ photolysis on planets around K and M  stars \citep{Selsis:2002,Tian:2014,Domagal:2014,Gao:2015,Harman:2015}. A third major method for producing large quantities up to 300~bars of O$_2$ is XUV-driven escape of H produced by photolysis of water vapor in a massive early steam atmosphere \citep{Luger:2015}. However, reactions with existing magma oceans or subsequent solid crustal layers can draw down O$_2$ to lower levels of a few bars \citep{Schaefer:2016,Wordsworth:2018,Barth:2021,Amaral:2021}.

None of these previous works have fully explored the time-dependent spectral variability of flares and their temperatures \cite[][and references therein]{Howard:2025}. Flares are generally assumed to distribute their energy at ultraviolet (UV) to optical wavelengths according to a 9,000-10,000~K blackbody when incorporated as stellar inputs for modeling exoplanet atmospheric photochemistry (e.g. \citealt{Segura:2010, Venot:2016, Tilley:2019, Chen:2021, Louca:2023}). Multiwavelength observations, however, support a wider range of flare effective temperatures spanning $\sim$4,000--40,000~K \citep{Kowalski:2013, Loyd:2018a, Froning:2019, Howard:2020, MacGregor:2021, Brasseur:2023, Jackman:2023, Paudel:2024, Corbett:2025, Howard:2025L, Kowalski:2025}. The ratio of far-UV to near-UV emission that sets photochemical conditions is 1700$\times$ higher for a 40,000 than a 4,000 K blackbody, underscoring the contribution of flares with different temperatures to the photochemical radiation environment \citep{Howard:2025}. Furthermore, flare-induced spectral variability in the stellar radiation environment depends on the combined effects of multiple flares in sequence or even flares that overlap in time. If an active star flares so frequently that an atmosphere does not have sufficient time to recover, it could lead to long-lasting depletion or enhancement of key biosignatures such as O$_2$ and O$_3$.  Ultimately, this could impact our ability to accurately assess habitability when we obtain snapshot observations of a planetary atmosphere.  With NASA's next flagship mission -- the Habitable Worlds Observatory (HWO) -- our field is undertaking the next steps in establishing whether the conditions for life exist elsewhere in the Universe \citep{Feinberg2026, Arney2026, Dressing2026}.  HWO architecture trade studies are actively underway with a launch likely in the 2040s.  Given this timeline, it is imperative that we predict and quantify expectations for living planets in these active systems now to understand our limitations and optimize our observing strategies to characterize the atmospheres of such worlds.

In this paper, we undertake the first study of how differences in the effective temperatures of repeated stellar flares impact the photochemical evolution of a young terrestrial planet atmosphere.  In Section \ref{sec:obsinputs} we describe observational constraints ton stellar flare rates that we use to obtain stellar spectral time series for input to the photochemical models described and presented in Section \ref{sec:model}. The results are presented in Section~\ref{sec:results}.  Section \ref{sec:discussion} discusses the implications of this work for future biosignature searches and the NASA Small Explorer (SMEX) mission concept EVE -- the Early eVolution Explorer.  Lastly, in Section \ref{sec:conclusions} we present our conclusions.

\section{Observational Stellar Flare Rates} 
\label{sec:obsinputs}

The dense cores of nearby clusters or associations of young stars do not suffer from significant membership confusion (i.e. which stars belong to which group) and have higher optical flare rates than those recorded from broader surveys of young stars \citep{Feinstein:2024, Howard:2025}. We previously measured average flare rates in the 0.6-1~$\mu$m TESS band for late K to mid M  stars in the dense cores of candidate EVE fields in \citet{Howard:2025}, which we now supplement with flare rates for young G star ($0.9<M_*<1.1$~M$_\odot$) and early K star ($0.7<M_*<0.9$~M$_\odot$) samples in the same fields to assess impacts for Earth analogs orbiting young solar-type stars. We identify G and K star flare-search targets using the larger EVE catalog \citep{Zhou:2026}, selecting all bright (TESS mag$<13.5$) G stars with ages of 10--30~Myr and early K stars with ages of 10--50~Myr ensuring sufficient numbers of high-photometric precision targets in the saturated flaring regime of ages for which occurrence rates remain relatively constant (e.g. \citealt{Davenport:2019,Ilin:2021,Feinstein:2024}). We then queried all 2 min TESS data on MAST and identified 1459.4 days of observations across 36 G stars and 1335.7 days of observations across 30 early K  stars, averaging $\sim$2 sectors of data per star.

Flares are identified with the \citet{Howard_MacGregor:2022} $\sigma$-cut flare-finder code as described for the late K and M-star sample in \citet{Howard:2025}. Briefly summarizing, the code pre-whitens the light curves to remove slowly varying astrophysical and instrumental signals with a Savitzky–Golay filter before searching for rapid flare signals. Candidates are identified in the cleaned light curves as sets of three adjacent points that exceed the local noise by 2.5$\sigma$ with at least one exceeding the noise by 3$\sigma$, and are each inspected by eye for a fast-rise, exponential decay profile to exclude false-positives from correlated noise or outliers. Flux amplitudes, start and stop times, and equivalent durations (EDs) are measured following \citet{Howard_MacGregor:2022} and flare energies measured by multiplying the EDs by the quiescent luminosities in the TESS band. In all, we identify 62 flares of 10$^{33.7-36.0}$ erg from the G  stars and 212 flares of 10$^{32.7-35.8}$ erg from the early K  stars.

\begin{figure}
    \centering
    \includegraphics[width=0.45\textwidth]{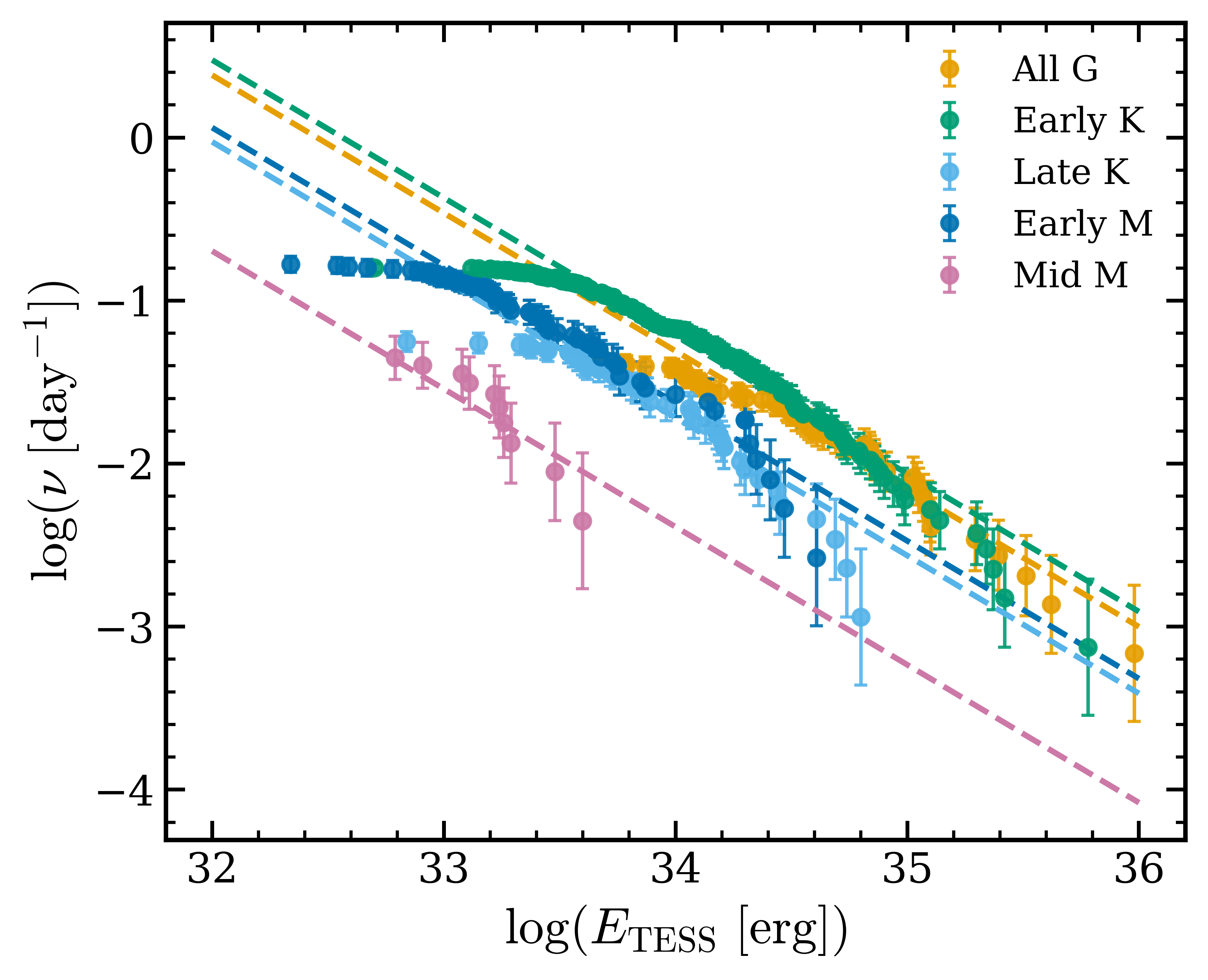}
    \caption{Flare frequency distributions (FFDs) measured from archival TESS data for G (orange), early K (green), late K (light blue), early M (dark blue), and mid M (purple) star stars show that all spectral types exhibit frequent flares at young ($<100$~Myr) ages.  Dashed lines indicate the best-fit power-law to each spectral type's FFD.}
    \label{fig:ffd}
\end{figure}

Next, we measure cumulative flare frequency distributions (FFDs) of the form $\log{\nu}=(1-\alpha)\log{E_\mathrm{flare}} + \beta$ averaged over all flares in the G and early K star samples, respectively (Figure~\ref{fig:ffd}). The number of flares d$^{-1}$ with energy $\geq E_\mathrm{flare}$ erg is given by $\nu$, the power law index determining the relative distribution of small and large flares is given by 1-$\alpha$, and the y-intercept describing the overall flare rate is given by $\beta$. We select all flares above the approximate detection completeness limits of 10$^{34.0-34.4}$ and 10$^{33.8-33.9}$ erg identified by the flattening of the cumulative rates of the G and early K samples, respectively. We then compute two iterations of FFDs each sample, leaving the power-law slope unbounded in the first iteration and reporting FFDs of log $\nu_\mathrm{G,free}$ = -0.66$\pm$0.04 log $E_\mathrm{TESS}$ + 21.1$\pm$1.30 and $\nu_\mathrm{K0-K5,free}$ = -0.90$\pm$0.03 log $E_\mathrm{TESS}$ + 29.3$\pm$1.02. Next, we fix $\alpha \approx$0.84--0.85 to match the power-law slopes of the late K to mid-M-star samples of \citet{Howard:2025} and report FFDs of log $\nu_\mathrm{G,fix}$ = -0.85$\pm$0.09 log $E_\mathrm{TESS}$ + 27.5$\pm$2.95 and log $\nu_\mathrm{K,fix}$ = -0.85$\pm$0.02 log $E_\mathrm{TESS}$ + 27.5$\pm$0.72, respectively. Fixing the power-law slopes in each stellar mass bin prior to the decrease in flare rate with spin down/age enables extension to solar-type stars of the \citet{Howard:2025} 2D model grid of flare rates as functions of stellar mass and age given a sole dependence on $\beta$. We justify the addition of the G and early K FFDs to the grid given the overlap or near-overlap in uncertainty regions on $\alpha$ in our FFD fits.

\section{Photochemical Modeling Approach} 
\label{sec:model}

\subsection{Synthesized Spectral Timeseries}
\label{synthstellarinputs}

\begin{figure*}[t]
    \centering
    \includegraphics[width=0.9\textwidth]{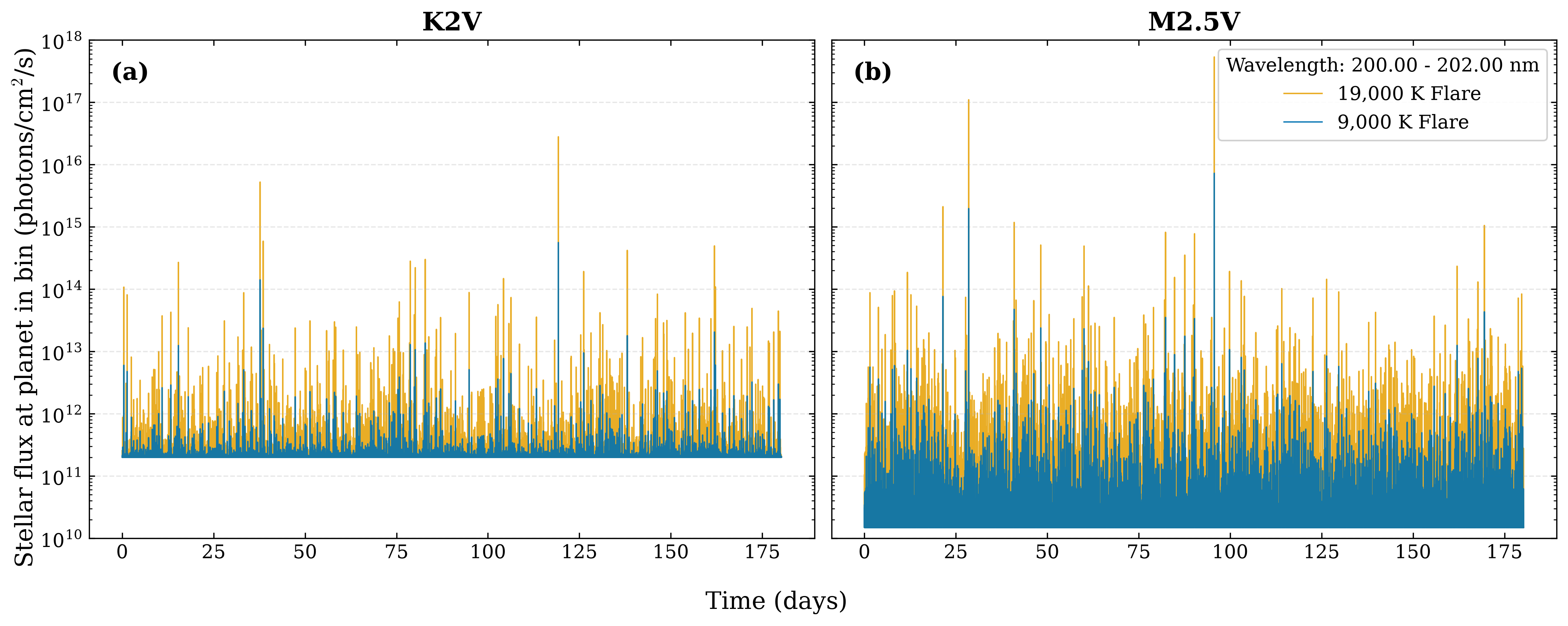}
    \caption{Stellar UV flux as a function of time for 180 days for a K2V star (left, panel (a)) and M2.5V (right, panel (b)) in the 201 nm bin due to a superposition of 26,380 flares (M2.5V) or 24,847 flares (K2V) sampled from the FFD, respectively. Flare energy varies as a function of wavelength governed by a 9,000~K and a 19,000 K blackbody, while the minimum flux represents the quiescent baseline.}
    \label{fig:lightcurves}
\end{figure*}

We choose $\epsilon$ Eri (K2V) and GJ 436 (M2.5V) as representative stars for potential HWO biosignature searches.  To generate 2D spectral time series as inputs for photochemical modeling, we start by sampling flares with $E_\mathrm{TESS}\geq$10$^{30}$~erg from the measured early K-star and M-star FFDs for a 180~d stare duration, scaling the TESS flare energies into the NUV using a constant scale factor of 0.475 for the 9,000~K scenario \citep{Howard:2025} and a NUV-TESS relation given by log $E_\mathrm{TESS}$ = 1.22 log $E_\mathrm{NUV}$ - 6.63 for the UV-luminous scenario \citep{Paudel:2024}. Next, we simulate NUV and optical time series for each flare at 30~s time resolution using the \citet{TovarMendoza:2022} flare template and inject the events into the 180~d time series at randomized times (Figure~\ref{fig:lightcurves}). Injected flares are drawn from an empirical power-law FFD and are injected randomly into the timeseries. The 30~s time resolution ensures that the flares are well-resolved. For $\epsilon$ Eri, we injected a total of 26,380 flares, and for GJ 436 we injected 24,847 flares. We set the randomly drawn flare injection times for the 9,000~K and 19,000~K scenarios to be the same so that only flare temperature is allowed to vary between the two scenarios.

\begin{figure*}
    \centering
    \includegraphics[width=0.9\textwidth]{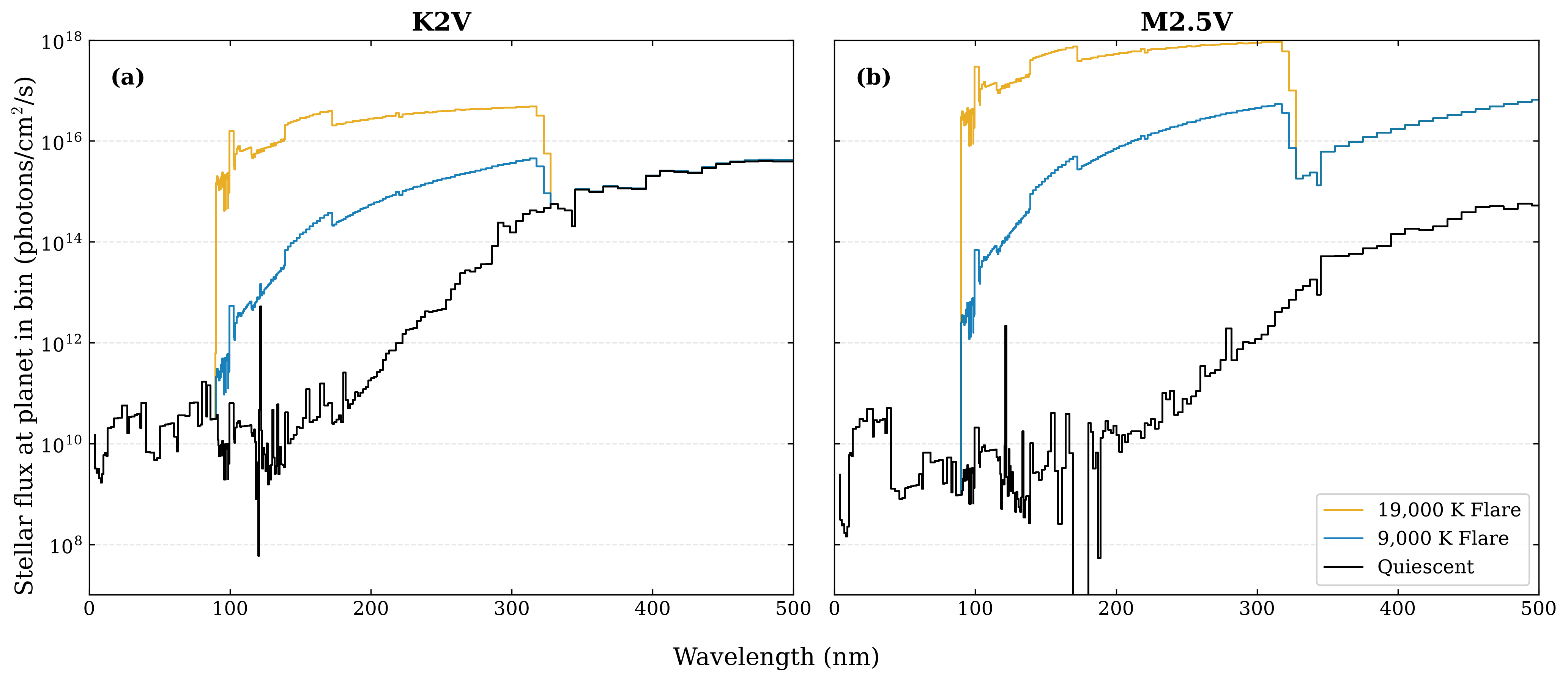}
    \caption{Stellar spectra received at the planet for a K2V star (left, a) and a M2.5V star (right, b). The black lines represent the baseline quiescent state of each star. The colored lines represent the flare spectra evaluated at their maximum 200 nm emission for two different flaring temperatures 9,000 K (blue lines) and 19,000 K (orange lines).}
    \label{fig:spectra}
\end{figure*}

The TESS band light curves are obtained from the time integral of the NUV light curves following \citet{Howard:2025} and light curve model fit parameters in each band (e.g., flux amplitude and FWHM duration) solved for consistency with the injected NUV and TESS band energies. We also generate flare temperature time series at UV wavelengths by normalizing the NUV light curves to a maximum temperature of either 9,000~K or 19,000~K and baseline temperature of 2,000~K. The flare temperature time series are used to synthesize blackbody flare spectra from 0.09—0.32~$\mu$m at each time step, where the filling factors are solved for consistency with the injected NUV fluxes. The resulting blackbody spectrum at each time step is then superimposed on a broadband $\epsilon$ Eri or GJ 436 spectrum from the MUSCLES archive \citep{Loyd:2016} to generate the final stellar input file for the 9,000~K or UV-luminous scenario (Figure~\ref{fig:spectra}).

\subsection{Photochem Simulations}
\label{sec:photochem}

To predict how the composition of an atmosphere would be altered by flares, we use the one-dimensional (1-D) photochemical model contained in the \texttt{Photochem} software package \citep{Wogan2025}. The photochemical code evolves a system of partial differential equations describing how atmospheric gases and aerosols are influenced by chemical reactions, condensation/evaporation, escape and vertical transport \citep[Equation 1 in][]{Wogan2025}. Here, we use the chemical network and thermodynamics included with v0.8.4 of \texttt{Photochem} \citep{Photochem2026} for species composed of H, N, O, and C. We omit sulfur-bearing molecules because our focus is on O$_3$ photochemistry, where sulfur is not of first-order importance \citep{Seinfeld2016}. 

All simulations have surface boundary conditions for a Proterozoic Earth-like world with 0.01~bar of O$_2$. We have assumed a Proterozoic-like world 100~Myr after planet formation as an exploratory scenario, which we note does not align with the timeline of Earth's Proterozoic eon that began $\sim$2~Gyr after planet formation. Evolutionary models predict that the pre-main sequence phase of low mass stars could drive significant hydrogen escape from a steam-dominated atmosphere and leave behind atmospheric O$_2$ (e.g., \citealt{Luger:2015}). Such a world may share similarities with Earth's Proterozoic eon including higher concentrations of atmospheric O$_2$ and a more oxidized surface environment at an earlier stage in its history. These results motivate the possibility of multiple evolutionary pathways and timelines toward Proterozoic-like O$_2$ levels \cite[][and references therein]{Zahnle:2013, Luger:2015, Meadows:2017}.

We fix the surface N$_2$ and CO$_2$ pressures to 0.79~bar and $4.05 \times 10^{-4}$~bar, respectively. Each model assumes a large surface reservoir of H$_2$O (i.e., an ocean), and imposes a relative humidity of condensation of 40\%, roughly consistent with Earth's global average. Following the modern Earth template described in \citet{Wogan2025}, we inject CH$_4$ and NO into the atmosphere from the surface at $1.4 \times 10^{11}$ and $3 \times 10^{9}$ molecules cm$^{-2}$ s$^{-1}$, respectively. All other species have surface deposition velocities following Table 6 of \citet{Ranjan2023}. Models use temperature and eddy diffusion profiles for modern Earth described in \citet{Wogan2025}, and assume both profiles are fixed as a function of time.

To simulate a planet experiencing a flare, our procedure is as follows. First, given the model setup described above, and a quiescent UV spectrum for either a K2V or M2.5V star, we integrate the model to a steady-state. Next, we impose the time-evolving flaring spectrum described in Section \ref{synthstellarinputs}, and integrate the model forward in time for 180~d with the steady-state atmosphere as an initial condition. We fix the maximum timestep to 30~s to match the simulated flare evolution timescale, and permit \texttt{Photochem}'s integrator to adaptively choose smaller time steps as necessary.

\section{Results} 
\label{sec:results}

\subsection{$O_3$ Column Depth}
\label{sec:column}


\begin{figure*}
    \centering
    \includegraphics[width=0.95\textwidth]{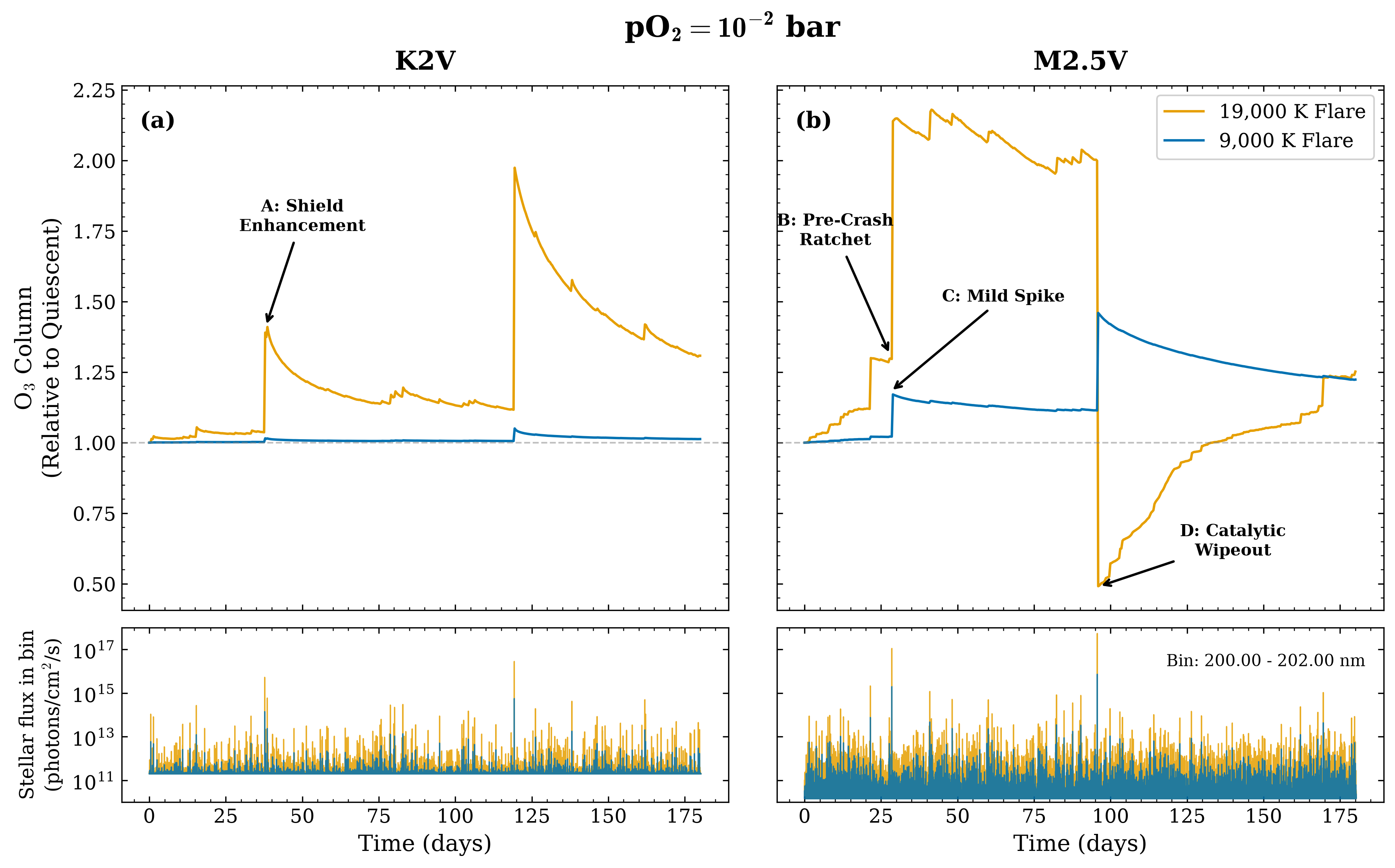}
    \caption{Time-dependent photochemical response of the stratospheric O$_3$ column for an Earth-like planet with a Proterozoic Earth-like atmosphere ($\rm pO_2 = 10^{-2}$ bar) orbiting a K2V star (left, panel (a)) and an M2.5V star (right, panel (b)). Top panels display the relative O$_3$ column abundance, with the quiescent baseline marked by the dashed gray line. Bottom panels show the driving stellar flux arriving at the top of the atmosphere in the 200-202 nm FUV bin. Responses to flaring sequences with equivalent blackbody temperatures of 9,000 K and 19,000 K are shown. While the K-star atmosphere exhibits smooth, isolated response peaks and exponential recovery curves following the flare events, the M-star atmosphere displays a stochastic behavior, including step-wise accumulation (``ratcheting") and sudden catastrophic depletion of O$_3$ around day 96. Labels A-D highlight four major events that describe the photochemical drivers of O$_3$ variability seen here: (A) Shield Enhancement at Day 38, (B) Pre-Crash Ratcheting at Day 28, (C) a Mild Spike at Day 28, and (D) a sudden Catalytic Wipeout triggered by an extreme flare at Day $\sim$96, which collapses the ozone shield to roughly half of its baseline value. Detailed column-integrated reaction rates for these photochemical events A--D are provided in Appendix \ref{sec:drivers} Figure \ref{fig:rxn_rates}.}
    \label{fig:Ozone}
\end{figure*}

Figure \ref{fig:Ozone} shows the temporal evolution of the O$_3$ column over a 180~d flaring sequence. The atmospheric response is heavily dependent on both the prescribed host star's spectral type and the flaring temperature. For the K2V star (Figure \ref{fig:Ozone}, panel a), the 9,000~K flares (blue) induce negligible changes to the O$_3$ column, while the hotter 19,000~K flares (orange) drive rapid enhancements of up to a factor of two relative to the quiescent baseline. The K-star planetary atmosphere exhibits smooth, exponential decay back toward the quiescent state following these events. Smaller flares briefly interrupt the exponential decay with small increases in O$_3$, while larger flares substantially increase O$_3$ and establish a new ``high water mark'' requiring several weeks to decay back down. In this way, we see that sequences of flares, particularly repeated strong flares, act together to ``ratchet'' up the O$_3$ column abundance over time as new flares occur before a full atmospheric recovery can take place. 

The M2.5V star (Figure \ref{fig:Ozone}, panel b) drives a stochastic photochemical environment. Since the quiescent UV baseline of the M-star case is lower, the high-frequency flares cause a more extreme version of the O$_3$ column upward ratcheting effect which is able to maintain an inflated state at ${\gtrsim}2\times$ the quiescent baseline for long periods. However, a major flare event at day ${\sim}96$ abruptly reverses this trend for the 19,000~K flare case, collapsing the O$_3$ column below the baseline to roughly 50\% of its quiescent value. Conversely, the 9,000~K flare case further increases O$_3$, similar to the flares in both the M- and K-star cases. Thus, the M-star atmosphere is highly volatile and capable of rapid shifts between net-production and net-destruction regimes.

Flare-driven increases in O$_3$ can be understood by analyzing the Chapman mechanism, which is contained within our chemical network (along with many other reactions):
\begin{align}
    \mathrm{O_2} + h\nu &\rightarrow 2\mathrm{O},
    & R_1 &= j_\mathrm{O_2}[\mathrm{O_2}], \\
    \mathrm{O} + \mathrm{O_2} + \mathrm{M}
        &\rightarrow \mathrm{O_3} + \mathrm{M},
    & R_2 &= k_2[\mathrm{O}][\mathrm{O_2}][\mathrm{M}], \\
    \mathrm{O_3} + h\nu &\rightarrow \mathrm{O_2} + \mathrm{O},
    & R_3 &= j_\mathrm{O_3}[\mathrm{O_3}], \\
    \mathrm{O} + \mathrm{O_3}
        &\rightarrow 2\mathrm{O_2},
    & R_4 &= k_4[\mathrm{O}][\mathrm{O_3}].
\end{align}
Here, $j_\mathrm{O_2}$ and $j_\mathrm{O_3}$ are the photolysis frequencies (units of 1/s), $k_2$ and $k_4$ are reaction rate coefficients, and species within brackets (e.g., $[\mathrm{O_2}]$) are number densities in molecules/cm$^3$. At a local steady state the four reactions above yield the following relationship between the O$_2$ and O$_3$ number density \citep{Seinfeld2016}:
\begin{equation} \label{eq:chapman}
    \alpha = \frac{[\mathrm{O_3}]}{[\mathrm{O_2}]}
    =
    \left(\frac{j_\mathrm{O_2} k_2[\mathrm{M}]}{j_\mathrm{O_3} k_4}\right)^{1/2}.
\end{equation}

We can build intuition for whether O$_3$ should increase or decrease by considering Equation \eqref{eq:chapman} for both a quiescent and flaring spectrum: $\alpha_\mathrm{quiet}$ and $\alpha_\mathrm{flare}$. The division $\alpha_\mathrm{flare}/\alpha_\mathrm{quiet}$ gives

\begin{equation} \label{eq:chapman_ratio}
    \frac{[\mathrm{O_3}]_\mathrm{flare}/[\mathrm{O_2}]_\mathrm{flare}}{[\mathrm{O_3}]_\mathrm{quiet}/[\mathrm{O_2}]_\mathrm{quiet}}
    \approx
    \left(\frac{j_\mathrm{O_2,flare}/j_\mathrm{O_3,flare}}{j_\mathrm{O_2,quiet}/j_\mathrm{O_3,quiet}}\right)^{1/2}.
\end{equation}
Here, $k_2$, $k_4$ and $[\mathrm{M}]$ are approximately independent of the incident spectrum, and so cancel. Furthermore, we can assume that $[\mathrm{O_2}]_\mathrm{quiet} \approx [\mathrm{O_2}]_\mathrm{flare}$ because in our simulations O$_2$ is a large reservoir, and so its number density is not greatly impacted by the flares relative to O$_3$. Consequently, Equation \eqref{eq:chapman_ratio} can simplify to

\begin{equation} \label{eq:o3_ratio}
    \frac{[\mathrm{O_3}]_\mathrm{flare}}{[\mathrm{O_3}]_\mathrm{quiet}}
    \approx
    \left(\frac{j_\mathrm{O_2,flare}/j_\mathrm{O_3,flare}}{j_\mathrm{O_2,quiet}/j_\mathrm{O_3,quiet}}\right)^{1/2}.
\end{equation}
Thus, simple Chapman chemistry predicts that flares would increase the O$_3$ abundance if the ratio of photolysis frequencies $j_\mathrm{O_2}/j_\mathrm{O_3}$ increases during a flare.  Figure~\ref{fig:rxn_rates} in Appendix~\ref{sec:drivers}, summarizes the integrated reaction rates for the most driving photochemical reactions at notable points in our simulations.

In Figure \ref{fig:Ozone}b, the 19,000~K flare case, we evaluate the photolysis frequencies at approximately 20\,km, within the stratospheric region where the Chapman O/O$_3$ chemistry is most relevant. The quiescent $j_\mathrm{O_2,quiet}/j_\mathrm{O_3,quiet} = 4.88\times10^{-9}$, while the $j_\mathrm{O_2,flare}/j_\mathrm{O_3,flare} = 6.22\times10^{-6}$ at 28.5 days, during the large flare that occurs there, corresponding to the large increase in O$_3$. Because O$_2$ and O$_3$ have different wavelength-dependent photolysis cross sections, the flare spectrum does not enhance their photolysis frequencies equally. In this case, the flare disproportionately increases the short-wavelength FUV flux (Figure \ref{fig:spectra}), producing a larger fractional increase in $j_\mathrm{O_2}$ than in $j_\mathrm{O_3}$ and therefore increasing their ratio.
Plugging these values into Equation \eqref{eq:o3_ratio}, we compute
$$\frac{[\mathrm{O_3}]_\mathrm{flare}}{[\mathrm{O_3}]_\mathrm{quiet}} \approx 36.$$
In practice, we do not see a 36-fold increase in the total ozone column (molecules/cm$^2$), because Equation \eqref{eq:o3_ratio} applies only to the number density of O$_3$ (in molecules/cm$^3$) at a discrete altitude. Furthermore, our photochemical simulation is time-evolving, and so the Chapman system does not necessarily have time to reach the Equation \eqref{eq:chapman} steady-state during a flare. Consequently, the total O$_3$ column increases by more moderate amounts (e.g., a factor of $\sim 2$ during a large flare). Although the Chapman mechanism is a simplified subset of the full photochemical network, it captures the essential shift in the O$_2$ and O$_3$ balance that explains the upward ratcheting of O$_3$ in our simulations.

However, in our simulations, the largest flares destroy O$_3$ rather than enhancing it, as is the case for the day-$\sim$96 flare in the 19,000~K M2.5V simulation (orange line in Figure \ref{fig:Ozone}b). During the peak of this large flare, direct O$_3$ photolysis overwhelms Chapman production. The Chapman formation rate $\mathrm{O}+\mathrm{O_2}+\mathrm{M}\rightarrow\mathrm{O_3}+\mathrm{M}$ is approximately $2\times10^{14}$ molecules cm$^{-2}$ s$^{-1}$, whereas the total O$_3$ photolysis loss rate reaches approximately $5\times10^{18}$ molecules cm$^{-2}$ s$^{-1}$. Thus,the immediate ozone collapse is driven primarily by the extreme photolytic loss of O$_3$.  The infographic in Figure~\ref{fig:infographic} in Appendix~\ref{sec:drivers} walks through each of the photochemical mechanisms that contribute to the dramatic O$_3$ depletion following this large flare event.

The same large flare simultaneously photolyzes H$_2$O, H$_2$O$_2$, HO$_2$, and CH$_4$, producing a large transient H/OH/HO$_2$ radical pulse. These radicals then accelerate the oxidation of methane and carbon monoxide toward H$_2$O and CO$_2$. For example, the coupled reactions
\begin{align}
    \mathrm{H}+\mathrm{O_3} &\rightarrow \mathrm{OH}+\mathrm{O_2},\\
    \mathrm{CO}+\mathrm{OH} &\rightarrow \mathrm{CO_2}+\mathrm{H}
\end{align}
have the net effect
\begin{equation}
    \mathrm{CO}+\mathrm{O_3}\rightarrow\mathrm{CO_2}+\mathrm{O_2},
\end{equation}
Methane oxidation supplies additional carbon monoxide and water through the intermediate hydrocarbon chemistry, while reactions like $\mathrm{OH}+\mathrm{HO_2}\rightarrow\mathrm{H_2O}+\mathrm{O_2}$ redistribute the radical pool. The resulting conversion of reducing material into CO$_2$ and H$_2$O, together with the direct photolytic loss of O$_3$, leaves the atmosphere in a chemically altered state that recovers only slowly after the largest flare.

\subsection{Simulated Reflected-Light Spectra}
\label{sec:reflected_light_spectra}

\begin{figure*}
    \centering
    \includegraphics[width=0.95\textwidth]{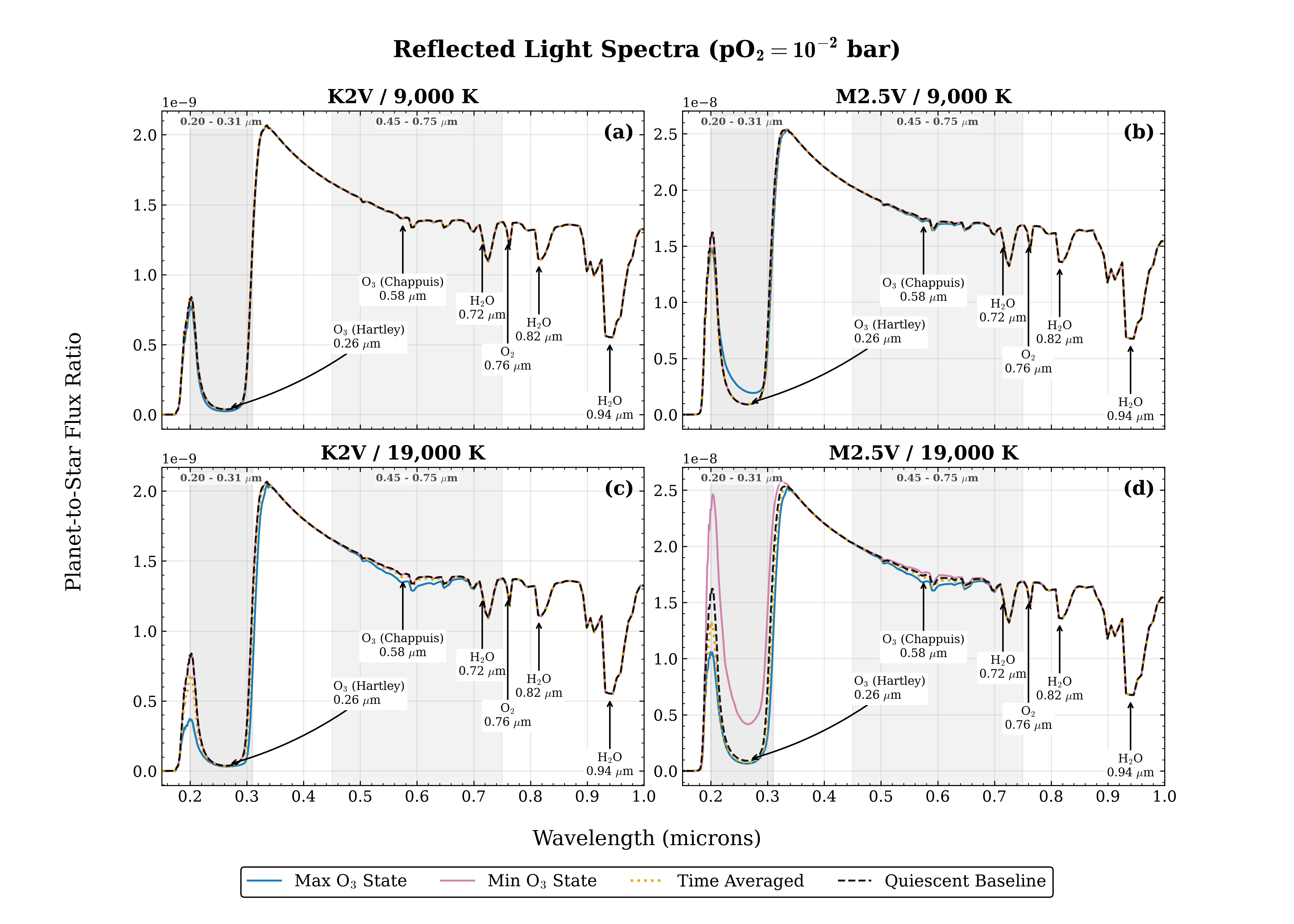}
    \caption{Simulated reflected light spectra (Planet-to-Star Flux Ratio) for a Proterozoic Earth-like exoplanet orbiting a K2V star (left; panels a,c ) and an M2.5V star (right; panels b, d) with two different flaring temperature hypotheses prescribed: 9,000 K (top row) and 19,000 K (bottom row). Spectra are calculated using \texttt{PICASO} at each extreme of the photochemical evolution: the quiescent baseline (dashed black), the composition of ozone averaged over time (dotted orange), the ozone column  at its maximum (max O$_3$) state (solid blue), and the ozone column at its minimum (min O$_3$) state (solid magenta). Key molecular absorption features are annotated for clarity. Shaded regions denote the physical extent of the primary ozone absorption features, including the primary UV Hartley (0.20--0.31 $\mu$m) and visible Chappuis (0.45--0.75 $\mu$m) bands. The near-infrared bulk atmospheric features for oxygen (0.76 $\mu$m) and water (0.72, 0.82, and 0.94 $\mu$m) are present. The UV and visible O$_3$ bands show a dependence on the stellar flare temperature: ozone production for K-star flares and ozone destruction for M-star flares. }
    \label{fig:PICASO}
\end{figure*}

To quantify the observational consequences of flare-driven photochemistry, we simulated the planet reflected light spectra of the $\rm pO_2 = 0.01$~bar atmospheres using the \texttt{PICASO} \citep{Batalha_2019, batalha_2026} radiative transfer model (Figure \ref{fig:PICASO}). The modeled wavelength regime (0.15-1.0 $\mu$m) captures key biosignature and habitability diagnostic features, including the O$_3$ Hartley and Chappuis bands, the O$_2$ A-band (0.76 $\mu$m), and H2O bands (0.72, 0.82, and 0.94 $\mu$m). Increases in O$_3$ throughout our simulations tend to result in broadening of the UV Hartley band around 0.3 $\mu$m and deepening of the Chappuis band in the optical.   

The HWO target star list includes a large number of K-stars where a $\sim$50\% change in the NUV reflected light spectrum between a 9,000 and 19,000~K flare sequence may be significant \citep{Tuchow:2025}. This promising diagnostic seen in our K-star models is most observationally apparent short of 0.25 $\mu$m in the blue wings of the Hartley band. Thus, access to these short wavelengths may be key for the HWO coronagraph to characterize the impact of flaring on ozone.    
For the M-dwarfs in the target star list \citep{Tuchow:2025}, HWO may only be sensitive to the UV because longer wavelengths are likely to be inaccessible due to inner working angle constraints in such compact systems. Moreover, our findings suggest that Earth-like planets around M-stars can exhibit large changes to ozone in the UV due to repeated flaring, making this a critical factor to consider for HWO observations of Earth-like planets around late-type stars.  
In particular, the drop in O$_3$ abundance from the largest 19,000 K M-star flare decreases the absorption in the core of the UV Hartley band by a factor of ${\sim}2\times$. UV retrieval estimates from \citet{Krissansen-Totton2026} suggest that such a drop in O$_3$ may be detectable provided that HWO has a UV coronagraph with a shortwave cutoff at or below 0.25 $\mu$m.

\section{Discussion}
\label{sec:discussion}

\subsection{The Role of O$_3$ in Planetary Habitability}
\label{sec:origins}

The presence or lack of O$_3$ can significantly impact the habitability of a planet since it is a key UV shield for surface life \citep{Segura:2003, Rugheimer_2015, tilley_kaltenegger_2019, Kozakis_2020}. Our models suggest different O$_3$ shielding due to flares.  These temporal dynamics observed at oxygen $\rm pO_2 = 0.01$~bar highlight a potential divergence in the habitability and observability of K-star versus M-star planets during Proterozoic-like eons. The K-star environment (Figure \ref{fig:Ozone}, panel (a)) provides a highly stable photochemical baseline. Since the star continuously emits a robust background of UV radiation, the atmosphere readily absorbs flare-induced perturbations, resulting in a predictable, temporary thickening of the primary UV shield, followed by a gradual return toward quiescent state. In contrast, the M-star environment (Figure \ref{fig:Ozone}, panel (b)) is characterized by a photochemical whiplash. The phenomenon that we term ``atmospheric ratcheting'' indicates that the planet rarely, if ever, exists in its modeled quiescent steady-state. Frequent small flares ratchet up O$_3$ levels, but the catastrophic O$_3$ depletion observed at day ~95 shows a potential evolutionary bottleneck; even an atmosphere with 1\% present-day levels of oxygen can have its UV shield temporarily cut in half by a single mega-flare. Thus, the combination of frequent small flares ratcheting up O$_3$ and infrequent large flares collapsing it back down produces a compelling dependence on the stellar FFD characteristics. The swings between O$_3$ extremes and the arrival at any new quasi steady-state are highly dependent on the relative frequency and temperature of the flares, which could be highly star-dependent, therefore motivating stellar observations for HWO target stars.   

The extreme temporal fluctuations seen in our model provide useful insight for future direct imaging missions. As seen in Figure \ref{fig:PICASO}, the maximum and minimum O$_3$ states depicted in Figure \ref{fig:Ozone} establish the absolute observational envelope of the planet over the flare sequences considered in our study. For the M-star, an observer could detect a deep Hartley absorption band during the ratcheted production phase, or reduced absorption features, following a depletion event which emphasizes a call to our communities that we need time-resolved, multi-epoch observations to accurately characterize M-star biospheres in the future. 

\subsection{The Early eVolution Explorer}
\label{sec:eve}

The Early eVolution Explorer (EVE) is a NASA Small Explorer (SMEX) mission concept \citep{MacGregor:2025} that will be submitted to the open Announcement of Opportunity (AO) in September 2026.  EVE is designed to study the early evolution of planetary systems, specifically the interplay between young stars, their surrounding protoplanetary disks, and their forming planets.  To achieve this science, EVE will observe simultaneously in three bands -- near-ultraviolet (NUV, 200--300~nm), visible (500--900~nm), and near-infrared (NIR, 1.0--1.6~$\mu$m.  A wide field of view (25 deg$^2$) and fast cadence (60~s) enable time series photometry of thousands of stars in young ($<100$~Myr) clusters.  EVE will use a 1200~km low-Earth orbit (LEO) providing sustained access to two 25$^\circ$ radius fields of regard centered at $0^\circ$ declination and $103^\circ$ and $283^\circ$ right ascension.  Every target cluster will be observed for a minimum of 45~days.  Additional details on target cluster selection are available in \cite{Zhou:2026}.

Although individual flares have been recorded with effective temperatures of 4,000--40,000~K as discussed earlier, we lack population statistics on flare temperature and how often flares of different temperatures occur. Only simultaneous, multi-wavelength observations can pin down the true temperature and energy of flares. Capturing the NUV band is especially critical given its importance to various photochemical processes (e.g., O$_2$, O$_3$, and SO$_2$). To fully understand the photochemical impact of flares, a much larger sample of energetic flares measured simultaneously in multiple bands with broad wavelength coverage is required. The only multi-wavelength studies of flares to date are the result of targeted monitoring campaigns \cite[e.g.,][]{MacGregor:2021,Jackman:2021,Tristan:2023,Paudel:2024}.  Only a handful of superflares have been observed simultaneously in two of the EVE bands (visible and NUV or NIR). No superflares have been observed in all three bands simultaneously. Long waiting times and unpredictable flare occurrence rates mean that coordinated campaigns between multiple observatories are challenging.  

A dedicated multi-wavelength mission is the only way forward to progress the state of the art. EVE’s novel three-band design will open a new window on flares, enabling more precise determinations of fundamental properties such as temperature and energy for the first large sample of events. These data will give us critical new insights into the photochemical processes that we have explored in this paper. Rather than simulating template time series data for representative stars, EVE will provide long-term monitoring of thousands of stars. This is a key piece of missing knowledge required for us to prioritize which stars and planetary systems to target with HWO. We also anticipate that EVE's multiband observations of both flares and planetary transits could improve constraints on the role of atmospheric escape in shaping young worlds (see Appendix~\ref{sec:escape} for further discussion).

\section{Conclusions}
\label{sec:conclusions}

In this paper, we present a new analysis of ozone photochemistry in a Proterozoic Earth-like atmosphere exposed to a 180-day time series of stellar flares.  We explore two different stellar types (K2V and M2.5V) and two stellar flare temperatures (9,000~K and 19,000~K) to bound the parameters space and explore which factors play the largest role in determining the potential habitability of the planet. Our main conclusions are as follows:

\begin{enumerate}
    \item Hot (19,000~K) flares make a more significant impact on ozone photochemistry for both the K2V and M2.5V star.  For the K star, hot flares drive rapid enhancements of the O$_3$ column up to 100\% relative to the quiescent baseline followed by smooth, exponential decay.  For the M star, smaller flares of both temperatures (9,000~K and 19,000~K) produce enhancements in the O$_3$ column, while a large 19,000~K flare collapses the O$_3$ column to roughly 50\% of its quiescent value.  
    \item M-star planetary atmospheres are highly volatile and can exhibit rapid shifts between net O$_3$ production and destruction.  Planets orbiting these late-type stars likely never exist in a truly quiescent state.  However, K-star planetary atmospheres more readily absorb flare-induced perturbations leading to greater stability.  These results emphasize previous suggestions that K stars might offer an advantage to future biosignature searches \citep[e.g.,][]{Heller_Armstrong:2014, Arney2019, Richey-Yowell:2022}. 
    \item Ultimately, all of our simulations rely on a fundamental understanding of flare rates and properties (energies and temperatures) that we currently lack on a population level.  If selected, the EVE SMEX mission concept would fill this gap providing robust temperature determinations as inputs to photochemical models for at least 10$^3$ flares from young FGKM stars.  
\end{enumerate}

Our results suggest many avenues for future exploration. Here, we adopt an initial oxygen pressure of $0.01$~bar representative of Proterozoic Earth.  To obtain a more complete picture of habitability over a planet's entire history, we need to consider different initial oxygen abundances.  Furthermore, our M-star simulations indicate that after repeated flares the planet might approach a quasi-steady state ozone column depth.  Unfortunately, 180~days is not sufficient to confirm this result and longer simulations are required. Although we explored two different flare temperatures, future work must also explore a grid of temperatures spanning the range between these bounding values.  Taken as a whole, these models will provide a comprehensive picture of flare-induced ozone photochemistry and vital information on potential biosignature false positives.  Completing this work in the next two decades is critical to more fully understand the landscape of planetary habitability and its impact on biosignature searches before the launch of HWO.  

\section*{Acknowledgments}

We thank Amber Young, Shawn Domagal-Goldman, Giada Arney, Knicole Colón, and Allison Youngblood for helpful conversations surrounding this project. 

MAM, JSC, and NFW acknowledge support for part of this research from the National Aeronautics and Space Administration (NASA) Interdisciplinary Consortia for Astrobiology Research (ICAR) under award number 19-ICAR19\_2-0041.

NFW also acknowledges support through the NASA Postdoctoral Program.

WSH acknowledges funding support provided by NASA through the NASA Hubble Fellowship grant HST-HF2-51531 awarded by the Space Telescope Science Institute, which is operated by the Association of Universities for Research in Astronomy, Inc., for NASA, under contract NAS5-26555.

GTM acknowledges support from the National Science Foundation MPS-Ascend Postdoctoral Research Fellowship under grant No. 2402296.

JLY was supported in part by the Virtual Planetary Laboratory, a member of the NASA Nexus for Exoplanet System Science (NExSS), funded via the NASA Astrobiology Program grant no. 80NSSC26K0615. 

\software{\texttt{Photochem} \citep{Wogan2025,Photochem2026}, \texttt{PICASO} \citep{Batalha_2019, batalha_2026}, \texttt{Matplotlib} \citep{matplotlib}, \texttt{NumPy} \citep{numpy}}

\bibliography{bib.bib}{}
\bibliographystyle{aasjournalv7}

\appendix

\section{Photochemical Drivers of O$_3$ Variability}
\label{sec:drivers}

\begin{figure*}[htbp]
    \centering
    \includegraphics[angle=90, width=0.5\textwidth]{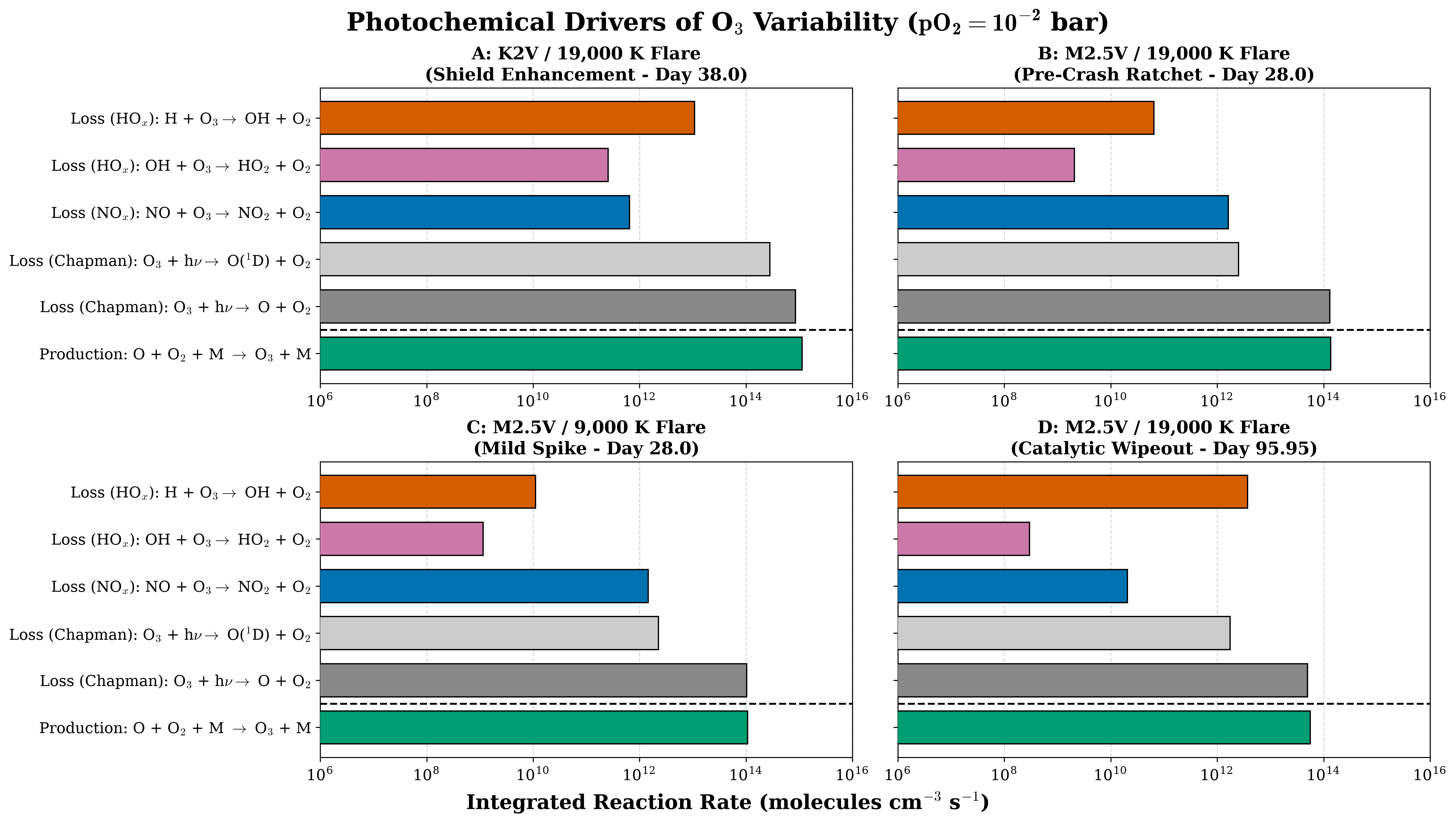}
    \caption{Visual summary of the column-integrated reaction rates for the dominant stratospheric O$_3$ production and loss pathways, extracted at the four key photochemical events (A--D) defined in Figure \ref{fig:Ozone}. The dashed black line in each panel physically delineates the Chapman production mechanism (seen on the bottom, bluish-green) from the primary chemical loss mechanisms (seen above the dashed black line). While Chapman photolytic loss pathways (grey bars) dominate the loss rates across all four events described here, they cancel each other out and create a null cycle that results in zero net loss of atomic oxygen and ozone. Catalytic destruction pathways driven by NO$_x$ and HO$_x$ radicals (highlighted in blue, purple, and vermillion) represent atomic oxygen and ozone sinks. Comparing the stable/enhancing events (panels A--C) to the Day 95.95 extreme flare event (panel D) reveals the kinetic mechanism behind the M-star ozone collapse: the extreme flare drives an HO$_x$-catalytic surge of several orders of magnitude, temporarily overwhelming Chapman production and pivoting the atmosphere into a severe net-loss regime.}
    \label{fig:rxn_rates}
\end{figure*}

In this appendix, we provide additional supporting materials to elucidate the photochemical drivers of ozone variability. 
Figure \ref{fig:rxn_rates} shows column-integrated reaction rates for the dominant stratospheric O$_3$ production and loss pathways that track the four key photochemical events (A--D) defined in Figure \ref{fig:Ozone}. 

\begin{figure*}[htbp]
    \centering
    \includegraphics[width=0.7\textwidth]{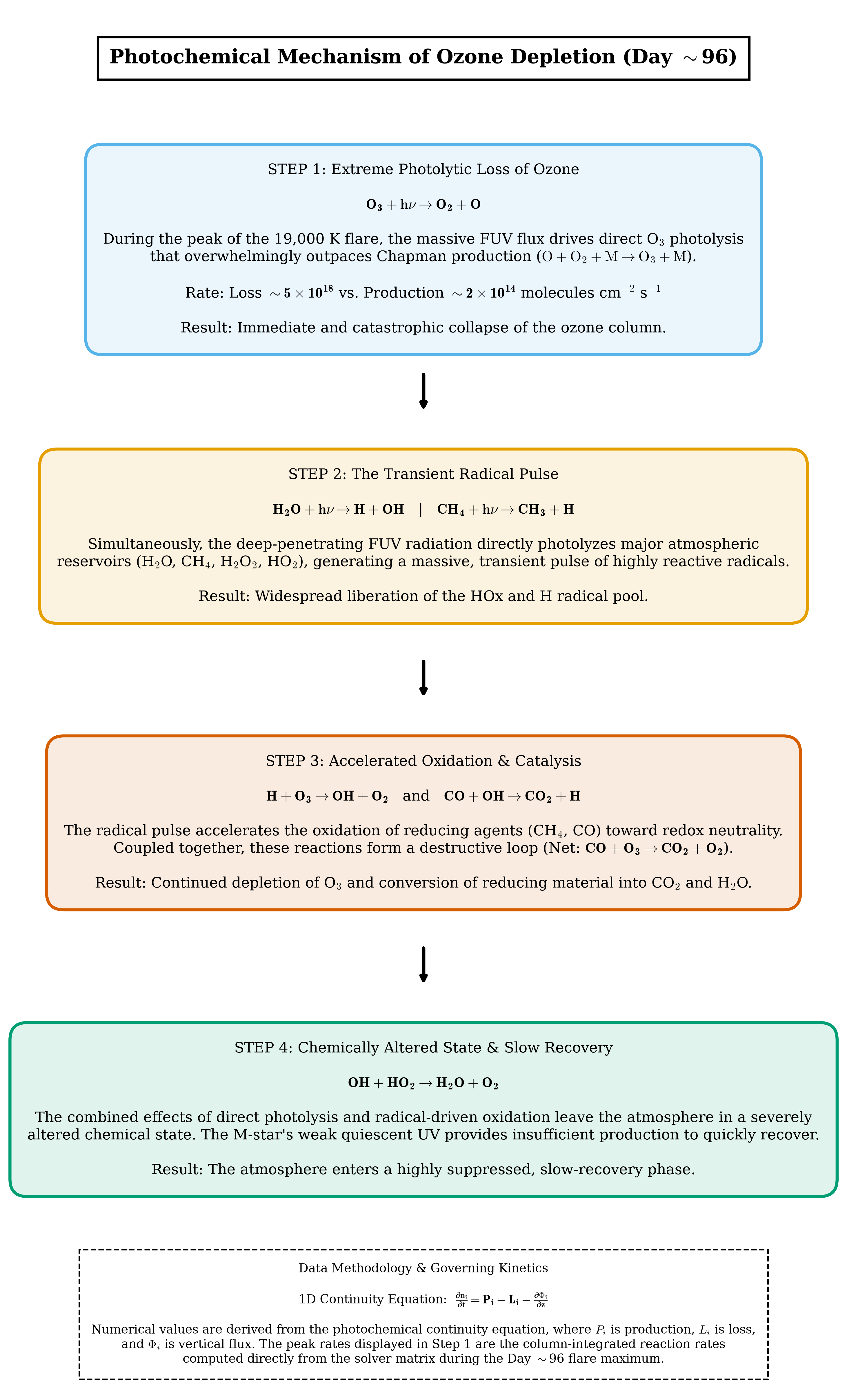}
    \caption{Step-by-step flow chart describing the photochemical mechanisms causing the ozone depletion seen on day $\sim$96 corresponding to the ``catalytic wipeout" described in Figures \ref{fig:Ozone} and \ref{fig:rxn_rates} for our 19,000 K M2.5V star case.}
    \label{fig:infographic}
\end{figure*}

Figure \ref{fig:infographic} shows shows a flow chart detailing the photochemical mechanisms that drive the ``catalytic wipeout" of ozone seen on day $\sim$96. This infographic summarizes the key details of the atmospheric transition from a sudden photolytically driven crash to a prolonged, catalytically suppressed state, via the following steps. Step 1: FUV radiation from a large flare directly photolyzes O$_3$ at rates that exceed the Chapman production rate. Step 2: FUV radiation from that same large flare simultaneously photolyzes H$_2$O, H$_2$O$_2$, HO$_2$, and CH$_4$, producing a large transient H/OH/HO$_2$ radical pulse. Step 3: These radicals accelerate the oxidation and reducing agents in the atmosphere, driving the system toward redox neutrality while also establishing a catalytic sink for atomic oxygen and ozone. Step 4: The M star's weak quiescent UV flux causes the surviving HO$_x$ radical pool to suppress ozone recovery. The direct photolytic loss of O$_3$, leaves the atmosphere in a chemically altered state that recovers only slowly after the largest flare. Displayed numerical values are vertically integrated column rates calculated directly from the 1D photochemical continuity equation at the Day $\sim$96 flare maximum.


\section{Atmospheric Escape}
\label{sec:escape}

Out of the full EVE flare sample, a subset of $\sim$600 will have $\geq5\sigma$ detections in all three bands, comprising a benchmark sample for model validation and multi-wavelength prediction. This sample will particularly benefit prediction of flare emission at the X-ray and extreme-UV (XUV) wavelengths that drive exoplanet atmospheric escape. Flares are expected to contribute at least 20--30\% of the total XUV fluence impacting planets around low mass stars \citep{Pass:2025,Howard:2025L}. However, most flare rate measurements from young stars have been obtained in a single optical band by facilities such as Kepler or TESS \cite[e.g.,][]{Davenport:2019,Ilin:2021,Froning:2019}. A broad grid of 6--40,000~\AA~flare spectra generated by RADYN beam-heating models show XUV counterparts consistent with a single broadband optical measurement are essentially unconstrained and cover a range of $10^8$ in allowed XUV flux densities \citep{Kowalski:2024}. 

We test the discriminatory power of NUV—NIR observations for our benchmark sample relative to single-band optical observations within the same RADYN model grid and demonstrate the addition of the NUV and NIR bands decreases the uncertainty in the XUV inference by four orders of magnitude. We derive the historic or cumulative optical flare energy emitted within the first 100~Myr for K and M  stars from TESS-band flare rates measured in young clusters \citep{Howard:2025} and propagate uncertainties of $10^8$ and $10^4$ in the optical-XUV flare energy relation for single and multi-band scenarios to compare with quiescent XUV emission, respectively. We find that adding NUV and NIR bands decreases uncertainty in the inferred historic XUV fluence by a factor of $\sim2$ due to the intermittent nature of flaring relative to quiescence. Atmospheric retention models show a factor of 2 variation in historic XUV fluence can mean the difference between atmospheric retention and loss for rocky planets near the cosmic shoreline that are considered high-priority targets for atmospheric characterization such as TRAPPIST-1 f and g or L 98-59 c and d \citep{Pass:2025}.

\end{document}